\documentclass[runningheads]{llncs}
\usepackage{graphicx}
\usepackage[inkscapelatex=false]{svg}
\usepackage{soul, color, color, makecell, multirow, bm}
\usepackage{tcolorbox}
\usepackage{adjustbox}
\usepackage{booktabs}
\usepackage{amssymb}
\usepackage{amsmath}
\usepackage{hyperref}
\usepackage{subcaption} % Required for creating subfigures

\begin{document}


%
       % typeset the header of the contribution
%
% \begin{abstract}
% The abstract should briefly summarize the contents of the paper in
% 15--250 words.

% \keywords{First keyword  \and Second keyword \and Another keyword.}
% \end{abstract}
%
%
%

\title{Unlocking the Potential of Large Language Models for Explainable Recommendations}


\author{
Yucong Luo \and
Mingyue Cheng \and
Hao Zhang \and
Junyu Lu \and
Qi Liu \and
Enhong Chen\
}
\authorrunning{Y. Luo et al.}
\institute{University of Science and Technology of China, Hefei, China
\email{\{prime666,zh2001,lujunyu\}@mail.ustc.edu.cn},
\email{\{mycheng, qiliuql, cheneh\}@ustc.edu.cn}
}

\maketitle       
\begin{abstract}

Generating user-friendly explanations regarding why an item is recommended has become increasingly prevalent, largely due to advances in language generation technology, which can enhance user trust and facilitate more informed decision-making during online consumption. However, existing explainable recommendation systems focus on using small-size language models. It remains uncertain what impact replacing the explanation generator with the recently emerging large language models (LLMs) would have. Can we expect unprecedented results?  In this study, we propose LLMXRec, a simple yet effective two-stage explainable recommendation framework aimed at further boosting the explanation quality by employing LLMs. Unlike most existing LLM-based recommendation works, a key characteristic of LLMXRec is its emphasis on the close collaboration between previous recommender models and LLM-based explanation generators. Specifically, by adopting several key fine-tuning techniques, including parameter-efficient instructing tuning and personalized prompt techniques, controllable and fluent explanations can be well generated to achieve the goal of explanation recommendation. Most notably, we provide three different perspectives to evaluate the effectiveness of the explanations. Finally, we conduct extensive experiments over several benchmark recommender models and publicly available datasets. The experimental results not only yield positive results in terms of effectiveness and efficiency but also uncover some previously unknown outcomes. To facilitate further explorations in this area, the full code and detailed original results are open-sourced at \href{https://github.com/GodFire66666/LLM_rec_explanation}{https://github.com/GodFire66666/LLM\_rec\_explanation-7028/}.
\end{abstract}

\section{Introduction}
Recommender systems \cite{resnick1997recommender} play a pivotal role in aiding users to navigate through enormous online content, personalizing recommendations based on user preferences \cite{cheng2022towards,cheng2021learning}. As recommender systems deeply impact personal decisions, it is increasingly crucial to ensure users' understanding and trust. This has led to the emergence of explainable recommendation models \cite{zhang2020explainable}, which not only predict users' preferences but also provide explanations for their recommendations. These explanations, serving as a bridge between users and the system, can enhance user trust, assist in better decision-making, and persuade users to try or buy an item. However, striking a balance between the accuracy of recommendations and the explainability of the models remains a challenge.

% \begin{figure}[h]
%   \centering
%   \includesvg[width=0.83\linewidth]{image/intro_case}
%   \vspace{-0.2in}
%   \caption{Explanations provided by different methods}
%   \vspace{-0.3in}
%   \label{fig:1}
% \end{figure}
Current explainable recommendation methods can be divided into two groups: Embedded methods and post-hoc methods. Embedded methods \cite{mcauley2013hidden,zhang2014explicit,diao2014jointly} integrate explanation directly into the recommendation model construction, using text  \cite{chen2018neural} or images \cite{chen2018visually} from item side information. However, their readability, consistency, and alignment with ratings are challenging, especially since explainability isn't their optimization goal \cite{wang2018reinforcement}. These retrieval-based methods may fail to provide personalized explanations in sparse data and face legal issues like copyright. Also, they require unique explanation strategies for different models.
Post-hoc methods \cite{bilgic2005explaining,tintarev2010designing,sharma2013social} aim to explain black-box models after training, often separating accuracy and explainability. They generate explanations from model outputs and pre-set templates like "people also bought". These are readable and persuasive but may not fully leverage the model's information or operational mechanism. Furthermore, the explanation diversity is limited by the number of pre-defined templates, possibly misrepresenting the model's actual reasoning. P5\cite{geng2022recommendation} and M6-Rec\cite{cui2022m6} employ language models for various recommendation tasks, including explainability, but they rely on user-item reviews as ground truth and utilize language models with a smaller number of parameters.

Recent advancements in large language models (LLMs) like those proficient in reasoning, knowledge utilization, and task generalization \cite{brown2020language}, suggest they could serve as effective zero-shot task solvers \cite{radford2019language}, trained on vast datasets for diverse applications \cite{gao2023chat,brown2020language,wei2022chain}. Despite their potential in feature extraction and text generation, integrating LLMs into recommendation systems to enhance explainability without sacrificing recommendation accuracy is challenging. Current research indicates a potential misalignment between recommendation effectiveness and explainability \cite{wang2018reinforcement}, and while LLMs may not be inherently suited for recommendations compared to traditional collaborative filtering, their advanced semantic understanding could improve the transparency and comprehensibility of recommendations. Our approach aims to leverage these capabilities of LLMs while preserving the core functionality of recommendation systems.

In this work, we propose a two-stage framework LLMXRec for post-hoc explainable recommendations. Essentially, this is a method for decoupling the recommendation model from the explanation generator, bearing resemblance to a post-hoc approach. The first stage is devoted to the training of the recommendation model, while the second stage concentrates on explanation generation. This framework is notably flexible, ensuring that the explanation generator neither compromises the accuracy of the recommendation model nor imposes restrictions on the type of recommendation models it can accommodate. We innovatively integrate LLMs as explanation generators, exploiting their potent capabilities in feature extraction, generalized reasoning, and text generation. However, it is crucial to point out that traditional template-based explainability evaluation metrics are not suitable for this context, thus necessitating the need to evaluate explanation quality. This overlooked aspect in prior studies is addressed in our work, aiming to deepen understanding of explainable recommendations.

The major contributions of this paper are:

\begin{itemize}
\item
We advocate for a two-stage framework LLMXRec that decouples item recommendation from explanation generation. Stage 1 can incorporate some current recommender models, offering the framework high flexibility and making it universally applicable. In stage 2, we use LLMs as explanation generators, leveraging their strong text understanding, knowledge consolidation, and inference capabilities to generate personalized textual explanations. 

\item
We utilize instruction tuning to enhance the precision and control of LLM-generated explanations. We've created a collection of high-quality explainable instruction datasets for tuning. Through instruction tuning, We bolster the higher controllability of LLMs, elevating the quality of the explanations they generate.

\item
We assess the efficacy of our approach by employing offline experiments of three aspects, covering both quantitative and qualitative measures. The results show our framework's effectiveness in interchanging various recommendation systems and explanation generators, with our instruction-tuned LLMs generating more useful explanations than the baseline.  The case study further illustrates our method's superiority.
\end{itemize}

\section{Preliminary}
\subsection{Problem Formulation.}
In this paper, we focus on post-hoc explanation generation. We define our problem as follows.

\textbf{Input}:
The input data of our model is the user set $U$, the item set $V$, the user-item interaction history $H$, a recommendation model $f$ that has been trained on $H$, and an explanation generator $G$. Given a user \bm{$u$} and his historical interaction records \bm{$h_u$}, the function $f(\bm{u})$ predicts top $k$ items that \bm{$u$} interests in as the recommended items. In the context of this study, $k$ is set to 1, denoted as the candidate item \bm{$v$}. The user \bm{$u$} from the set $U$ includes user ID and some side information (e.g., age and gender). The item \bm{$v$} from the set $V$ includes item ID and some features (e.g., category).

\textbf{Output:}
Given a user \bm{$u$} and their historical interaction records \bm{$h_u$}, we obtain a recommended candidate item \bm{$v$} through the recommendation system $f(\bm{u})$. Subsequently, the explanation generator $G$ provides a natural language explanation $Z$ as to why the recommendation system would suggest item \bm{$v$} to user \bm{$u$}, based on the information from \bm{$u$}, \bm{$h_u$}, and \bm{$v$}. The final output of this process is the natural language explanation $Z$.

\subsection{Instruction Tuning.}
The process of instruction tuning is an essential part of training Large Language Models (LLMs). This prevalent method enhances the competence of LLMs by tailoring a broad range of human-labeled instructions and responses [33]. This phase bestows the model with a strong capability for generalization, thus equipping it to adeptly handle new tasks and guaranteeing its flexibility in the face of unfamiliar situations and problems. Specifically, the process of instruction tuning encompasses four primary phases:
\begin{itemize}
\vspace{-0.05in}
\item Phase 1: Identify a particular task and create a clear, easily comprehensible instruction to address it effectively. The instruction should offer a detailed description of the task and appropriate strategies for managing it.
\item  Phase 2: Construct inputs and outputs for the task.
\item  Phase 3: Combine the instruction and input using natural language as a unified input for the entire model for each data point as "instruction input".
\item  Phase 4: Fine-tune the LLMs utilizing the structured "instruction input" and "output" paired data points.
\end{itemize}

\section{METHODOLOGY}
\vspace{-0.3in}
\begin{figure*}[h]
  \centering
  \includegraphics[width=1\linewidth]{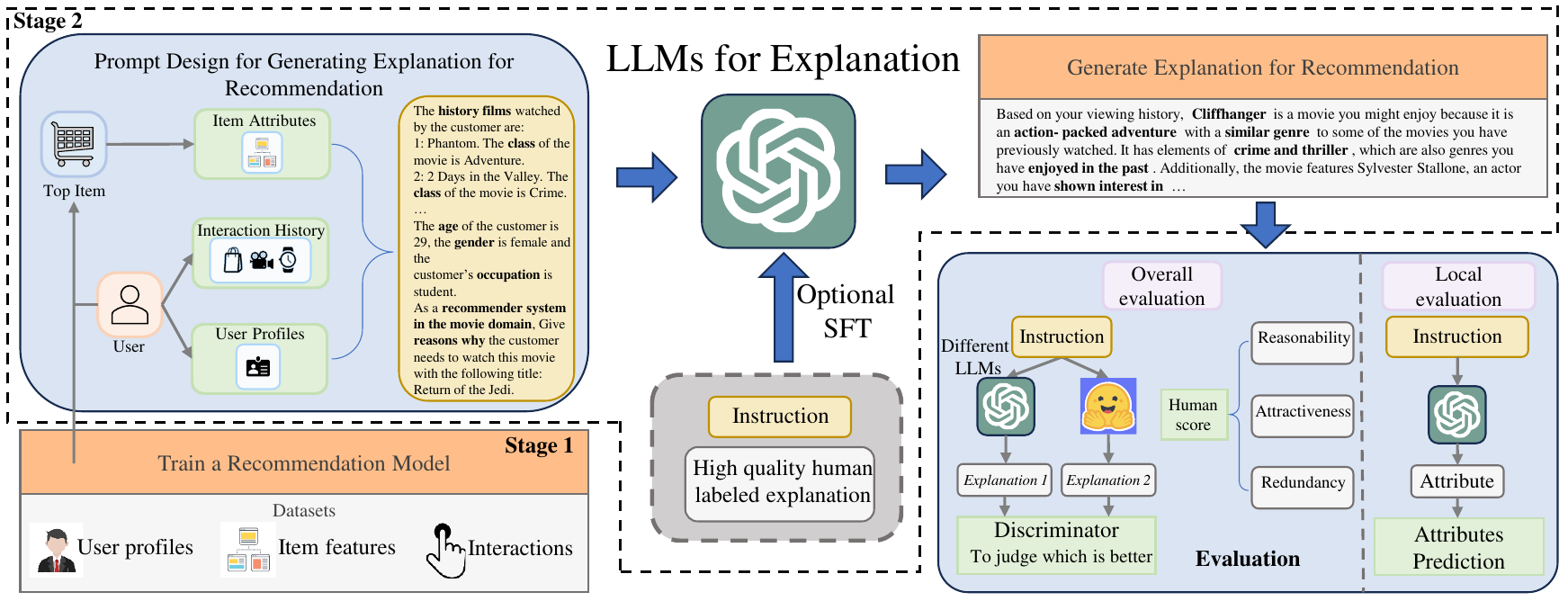}
  \vspace{-0.2in}
  \caption{The two-stage framework and evaluation}
  \vspace{-0.2in}
  \label{fig:2}
\end{figure*}
\vspace{-0.3in}

\subsection{Overview of the Two-stage Framework}
We provide an overview of LLMXRec framework, shown in \textbf{Figure \ref{fig:2}}. The proposed framework is decoupled into two distinct stages.

In the first stage, a recommender system, unrestricted by model type, is trained using user-item interaction data. Its core objective is to leverage this data to propose a recommendation item list for individual users.

In the second stage, an item \bm{$v$} is selected from the recommendation list for a specific user \bm{$u$}. The objective of stage 2 is to elucidate the reasoning behind recommending  \bm{$v$} to  \bm{$u$}. To generate explanations, LLMs are employed for natural language generation. Certain features may be inputted into the LLMs in textual form, such as the user's interaction history, user profile, item features, among others. Following this, the prompt is adjusted to stimulate the LLMs to furnish an explanation for the given recommendation item \bm{$v$}.

Following the generation of explanations, we propose three methods for assessing the quality of the explanations: For the overall quality assessment, (1) we train a discriminator to evaluate the relative quality of two explanations generated for the same user and recommendation item; (2) we employed human evaluation to rate the explanations. For the assessment of local explanation attributes, (3) we utilized a strategy wherein the explanation generator predicts attributes not fed into it, thereby assessing the explainability of local attributes.

\subsection{Explanable Generator Construction }
\subsubsection{Foundation Model Selection.}
\label{sect:3.3.1}

Our work primarily concentrates on the second stage, where we select LLMs as our explanation generators. The explanation generator is intended to yield comprehensible and precise explanations for the items recommended. Given LLMs' superior generalization, inferential skills, and fluency in language generation, and their performance in few-shot tasks, we adopte LLMs as the foundational models forexplanation generation. However, due to their resource of intensive nature - requiring significant hardware, time, and preprocessed data, we choose to use pre-trained LLMs instead of training new ones from scratch. Here we choose LLaMA\cite{touvron2023llama2} as backbone model.
% Leveraging LLMs as explanation generators, we utilize their prowess to produce explanations for recommendations, addressing the aforementioned tasks. Specifically, for each user \bm{$u$}, we first create a natural language template encapsulating their sequential historical interaction records $H_u$ and a candidate item \bm{$v$} from the recommendation list. These patterns are then incorporated into this template, forming the final instructions. Consequently, LLMs are able to interpret these instructions, generating comprehensible explanations consistent with the provided instructions.

% \definecolor{g}{RGB}{229, 239, 219}
% \definecolor{Emerald}{rgb}{0.31, 0.78, 0.47}
% \sethlcolor{g}

\subsubsection{Instruction Template Construction.}
\label{sect:3.3.2}
% Given that LLMs can only accept text as input, to enable LLMs to comprehend and execute the recommendation explanation task, we formulate the instructions into a natural language template and feed it into the LLMs.Here are the steps to construct a natural language template:

Since LLMs require text input, we convert the recommendation explanation task into a natural language template for LLM comprehension and execution. Here are the steps to create such a template:

\begin{itemize}

\item 
\textbf{System Instruction. }
To help LLMs grasp the explanation generation task, they need system instruction to set the working context, outlining input and output specifications. This aids LLMs in fully understanding their role as explanation generator. 
% In order to enable LLMs to comprehend the explanation generation task at hand, it is necessary to provide them with system instruction prior to presenting the detailed content. This instruction establishes the working context for the LLMs, delineating input and output specifications, and thereby facilitating a more comprehensive understanding of the task for the LLMs. 

\item 
\textbf{User’s Historical Interaction Records.}
Drawing inspiration from prior research, we sequence user interactions chronologically for LLM input. Additionally, we can incorporate user profiles and certain item features (e.g., category information) into the input of LLMs. We posit that adding extra side information enables LLMs to extract the relationship between users and items more comprehensively, thereby yielding more reliable explanations. 

\item 
\textbf{Chain of Thought (CoT).}
% CoT is a prominent prompting method for LLMs to tackle a variety of tasks. It is defined as a series of intermediate reasoning steps, aimed at enhancing the complex reasoning capabilities of LLMs. For the task of personalized recommendation explanation, we can incorporate guidance into the prompt, enabling the LLMs to generate more accurate explanations by emulating human thought processes. 
CoT, a key prompting technique for LLMs, involves intermediate reasoning steps to boost complex reasoning. For personalized recommendation explanations, guidance in the prompt lets LLMs mimic human thinking for more precise explanations.

\item 
\textbf{The Candidate Item.}
Upon incorporating user historical interaction information, we need to include information about the candidate items and require the LLMs to provide explanations.  
\end{itemize}

We integrate the above modules into a single instruction and feed it into the LLMs. For a specific user \bm{$u$}, we generate the top 1 recommended item \bm{$v$} through Stage 1. Subsequently, we have: 
\begin{equation}
I_{e} = T_e(\bm{u},\bm{i},\bm{h_{u}})
\label{eq:1}
\end{equation}
where $I_e$ denotes the instruction for explanation generation. $T_e$ is the instruction template for explanation generation. Following this, we obtain: $Z = G(I_e)$
, where $Z$ symbolizes the generated explanation, and $G$ signifies the explanation generator, i.e., LLMs. The template is constructed as follows:

\begin{tcolorbox}
% [colback=Emerald!10,colframe=cyan!40!black]
[title=\textbf{Instruction for Generating Explanation}]
\label{template:1}
\itshape
\textbf{The history films} watched by the customer are:
$<$HisItem Title and Class List$>$.
The \textbf{age} of the customer is $<$age$>$, the \textbf{gender} is $<$gender$>$ and the customer's \textbf{occupation} is $<$occupation$>$. As a recommender system in the movie domain, give \textbf{reasons why} the customer needs to watch this movie with the following \textbf{title and class}:

$<$Candidate Item Title and Class$>$
\end{tcolorbox}

\begin{tcolorbox}
% [colback=Emerald!10,colframe=cyan!40!black]
[title=\textbf{Instruction for Attribute Prediction}]
\label{template:2}
{\itshape
The history games played by the customer are:
$<$HisItem Title and Attribute List$>$. The recommender system suggests the customer to play this game with the following title:
$<$Candidate Item Title$>$.

\textbf{Your mission is to infer the game's information from the history record, such as $<$Attribute$>$.} You must infer $<$Attribute$>$. Do not return other information. And DO NOT return Unknow or Null.
}
\end{tcolorbox}

\subsubsection{Parameter-Efficient Instruction Tuning.}
\label{sect:3.3.3}
% Instruction tuning is crucial in managing uncontrolled outputs in LLMs because it allows for more nuanced control over the model's behavior, without which the model might produce outputs that are technically correct but not useful or understandable to the user.  By fine-tuning the model with specific instruction and output pairs, we can guide the model to produce outputs that align more closely with our desired explanations.  Specifically,
We utilize the conditional language modeling objective during the instruction tuning, as exemplified in the Alpaca \cite{Alpaca}. Formally,
\begin{equation}
    \max_{\Phi} \sum_{(x,y)\in M }\sum_{t=1}^{|y|}log(P_\Phi(y_t|x,y_{<t}))
\end{equation}
where $\Phi$ is the original parameters of LLMs,  $M$ is the training set, $x$ and $y$ represent the "instruction input" and "output" in the self-instruct data, respectively, 
$y_t$ is the $t$-th token of the y and $y_{<t}$ represents the tokens before $y_t$, 
% On the other hand, parameter-efficient tuning, such as Low Rank Adaptation (LoRA) \cite{hu2021lora}, is necessary to make the fine-tuning process more feasible and efficient. The standard fine-tuning procedure involves updating all the parameters of the model, which can be computationally expensive and time-consuming, especially for LLMs. However, by incorporating a low-rank matrix into the weight matrices of the model, LoRA allows us to update a smaller number of parameters while still achieving significant improvements in performance. This makes the fine-tuning process faster and less resource-intensive, which is particularly beneficial when dealing with LLMs or when computational resources are limited.

Efficient tuning is crucial for the viability of fine-tuning in LLMs, as traditional methods are resource-heavy. Noting that LLMs have surplus parameters, we use a more streamlined finetune approach, like Low Rank Adaptation (LoRA). By incorporating a low-rank matrix into the model's weights, LoRA allows fine-tuning a smaller parameter subset, achieving lightweight fine-tuning. This method improves performance and saves computational resources and time. Specifically, LoRA freezes the pre-trained parameters and adds trainable rank decomposition matrices to each Transformer layer, enabling efficient information integration, making fine-tuning faster and less resource-consuming, beneficial for LLMs in limited-resource situations. The final learning objective is computed as: 
\begin{equation}
    \max_{\Phi} \sum_{(x,y)\in M }\sum_{t=1}^{|y|}log(P_{\Phi+\theta}(y_t|x,y_{<t}))
\end{equation}
where $\theta$ is LoRA parameters and we only update $\theta$ during the training process.

\subsubsection{Instruction Tuning Data Construction.}

Given the imperative to optimize LLMs with specific instructions, we build a robust, high-quality instruction dataset for explanations. The high-quality dataset provides explicit, precise explanation, improving learning efficiency and reducing training resources. Furthermore, it bolsters the model's generalization and anticipation of unseen scenarios.

% Initially, we formulate these instructions adhering to the template delineated in  Equation \ref{eq:1}. To ensure the diversity of the instruction set, we generate diverse instructions by incorporating historical interaction \bm{$h_u$} of varying lengths $L$ and recommended items \bm{$v$} generated by different recommendation models $f$.  We then employ manual annotation to guarantee superior quality labels for them. This ensures that the labels are not only accurate and reasonable but also exhibit minimal redundancy. We also built an attribute prediction dataset using the templates from Equation \ref{eq:1}.  Instead of using them for explanation generation, we repurpose them for attribute prediction. The labels correspond to the ground truth attributes, encompassing user profiles, item features, and global attributes, such as popularity. We combine two datasets for multi-task instruction tuning, achieving both overall and local explanation.

Initially following Template \ref{template:1}, we craft varied instructions using user histories \bm{$h_u$} of different lengths $L$ and items \bm{$v$} from various models $f$. Manual annotation ensures the labels are accurate, reasonable, and not redundant. We also create an attribute prediction dataset using Template \ref{template:2}, targeting attribute prediction rather than explanations. Labels match true attributes, covering user profiles, item details, and global features like popularity. We merge these datasets for multi-task tuning, aiming for both overall and local explanations. After construction of this instruction dataset, we implement the efficient-parameter tuning method mentioned in Section \ref{sect:3.3.3}.

\subsection{Evaluation of Generated Explanations}

% Within the context of our two-stage framework, traditional evaluations don't apply due to lack of a ground truth. Given this constraint, our evaluation philosophy leads us to adopt an offline evaluation methodology, designed to assess the efficacy of various LLMs in generating recommendation explanations. This comprehensive assessment strategy provides a balanced view of text generation quality, taking into account both macroscopic and microscopic perspectives.

% Macroscopically, we train LLMs as discriminators to judge explanation quality using annotated data focus on ranking. We also employ human evaluations utilizing a scoring regression approach. Microscopically, we analyze individual attributes predicted by LLMs, adopting a classification strategy. 

In our two-stage framework, traditional evaluations aren't applicable due to the absence of ground truth. Hence, we use an offline evaluation method to assess LLMs' effectiveness in generating recommendation explanations, considering both macroscopic and microscopic perspectives.

Macroscopically, we train LLMs to discern explanation quality via ranked annotated data and implement human scoring regression evaluations. Microscopically, we examine LLM-predicted attributes using a classification approach.

\vspace{-0.1in}
\subsubsection{Automatic Evaluation with Fine-tuned LLMs as Discriminator.}
\label{sect:3.4.1}
% We have devised a new assessment mechanism: training a LLM to act as a Discriminator to discern the quality of the explanations.

% \textbf{Discriminative Instruction Data Generation.}
% Given the powerful semantic understanding capabilities of the LLMs, we propose instruction tuning the LLMs to serve as discriminators. It is acknowledged that the comparative ranking proficiency of LLMs surpasses their absolute ranking capability. Consequently, the training data we supply is designed to discern the relative quality of two explanations generated from the same instruction.  Formulary, 
% \begin{equation}
% D(T_d(I_e,Exp_1, Exp_2))=
% \left\{ \begin{array}{ll}
%              1, & \text{if $Exp_1$ is better} \\
%              2, & \text{if $Exp_2$ is better} \\
%            \end{array} \right.
% \label{eq:4}
% \end{equation}
% where $D$ denotes the discriminator and $T_d$ is the instruction generation function for discrimination. $I_e$ denotes the instruction used for generating explanations, and $Exp_1$ and $Exp_2$ refer to the explanations generated by different LLMs, respectively. 

Due to LLMs' semantic prowess, we propose instruction tuning to make them discriminators. LLMs excel at comparative over absolute quality ranking. We train them to judge the quality between two same-instruction explanations. Formally,
\begin{equation}
D(T_d(I_e,Exp_1, Exp_2))=
\left\{ \begin{array}{ll}
             1, & \text{if $Exp_1$ is superior} \\
             2, & \text{if $Exp_2$ is superior} \\
           \end{array} \right.
\label{eq:4}
\end{equation}
where $D$ is the discriminator, $T_d$ the discrimination instruction template, $I_e$ the explanation-generating instruction, and $Exp_1$, $Exp_2$ the different LLMs' explanations.

Given an $I_e$ and explanation pairs $(Exp_1, Exp_2)$ generated by different LLMs, we utilize both manual annotation and GPT4 for labeling. The annotation labels are either 1 or 2. To eliminate bias, we selected data points where the human annotation and GPT4 annotation annotate the same label as training data for the discriminator. Each data point consists of an instruction input generated by Equation \ref{eq:4}, and the annotated label as output.
% signifying the superior quality of either the first or second explanation respectively. This method enables explanation ranking based on quality, serving as training data for LLM fine-tuning.

Upon completion of the instruction data construction, we also employ LoRA to execute instruction tuning on the LLMs. Subsequent to this fine-tuning process, the LLMs attain the capability to function as a Discriminator, being able to discern the quality of two explanations derived from a single instruction. The template is constructed as follows:

\begin{tcolorbox}
% [colback=Emerald!10,colframe=cyan!40!black]
[title=\textbf{Instruction for Discriminator}]
{
\textbf{Instruction} 

{\itshape 
You are a discriminator that \textbf{judges whether the explainability} of the recommendation system is good or bad. You should judge \textbf{which of the 2 explainable opinions generated based on the following Instruction is better}. Return 1 if you think the first one is better, and 2 if you think the second one is better. 

\textbf{--------------------Instruction--------------------}

$<$Instruction for generating explanation$>$

\textbf{--------------------Explanation 1--------------------}

$<$Explanation 1$>$

\textbf{--------------------Explanation 2--------------------}

$<$Explanation 2$>$

\textbf{---------------------------------------------------------------}

Based on the above instructions, decide which explanation better explains why the recommendation system recommends this item to the customer. Please return 1 or 2 to show your choice. Only return 1 or 2.
}   

\textbf{Output} 
1
}
\end{tcolorbox}

\subsubsection{Manual Evaluation with Scoring Explanation.}
\label{sect:3.4.2}
To evaluate the efficacy and quality of the explanations, and to ascertain their actual utility to users in the context of recommendations, we utilize human evaluation. This evaluation is segmented into three distinct aspects for a comprehensive scoring methodology:

\begin{itemize}

\item 
\textbf{Reasonability}: The reasonableness of the explanation based on the instructions provided.

\item 
\textbf{Attractiveness}: The extent to which the explanation attracts users.

\item 
\textbf{Redundancy}: The degree of redundancy in explanation, which indicates whether users possess the patience to continue reading.
\end{itemize}
To ensure the objectivity and fairness of the evaluation process, we employ a blind scoring methodology. In this approach, human evaluators are not privy to the origin of the corresponding instruction interpretations, specifically, they are unaware of the LLMs from which these explanations have been generated. This method helps to eliminate potential bias and facilitates a more impartial assessment of each LLM's performance.

\subsubsection{Local Evaluation with Attribute Prediction.}
\label{sect:3.4.3}
% The Discriminator focuses on the ability to rank the quality of multiple explanations, while the focus of human evaluation is on scoring individual explanations. Both of them are evaluations of the overall explanation, yet we still require a method to assess the capacity for local explanation.
The Discriminator ranks multiple explanation qualities, whereas human evaluation scores single explanations. Both assess overall explanation quality, but a method for evaluating local explanation capacity is still needed.

For a user \bm{$u$}, we input historical interaction sequences \bm{$h_u$} and candidate items \bm{$v$} into the LLMs. Instead of generating comprehensive explanations, our focus pivots towards predicting specific local attributes that require to be explained, such as user profiles or item attributes.
Formulary,
\begin{equation}
    I_a = T_a(\bm{u},\bm{i},\bm{h_{u}})
\end{equation}
% where $I_a$ denotes the instruction for attribute prediction. $T_a$ denotes the instruction generation template for attribute prediction, which is shown in Section \ref{sect:3.3.2}. Then, $A = G(I_a)$,
% where A represents the predicted attribute.
% The capacity of the LLMs to explain local information is assessed by determining the accuracy of the predicted attributes.
\noindent
where $I_a$ is the attribute prediction instruction, and $T_a$ is the corresponding generation template in Section \ref{sect:3.3.3}. Thus, $A = G(I_a)$ signifies predicted attribute. We measure LLMs' local explanation ability by the accuracy of these attributes.

\section{Experiments}

In this section, we outline the experimental setup and evaluate the performance of various large language models (LLMs) in explanation generation and local feature prediction. We further explore means to improve explanation quality, concluding with a qualitative analysis of two explanation instances. We use LLMXRec to denote the instruction-tuned versions of LLaMA as explanation generator in subsequent sections.

\vspace{-0.35in}
\begin{table}
    \centering
    \caption{Dataset Information}
    \tabcolsep=0.1cm
    \renewcommand{\arraystretch}{1}
    % \vspace{-0.15in}
    \label{tab:1}
    \resizebox{1\columnwidth}{!}{%
        \begin{tabular}{cccccc}
        \hline
			\toprule
			Datasets & \multicolumn{1}{l}{\#Num. Users} & \multicolumn{1}{l}{\#Num. Items} & \multicolumn{1}{l}{\#Num. Actions} & \multicolumn{1}{l}{\#Actions/User} & \multicolumn{1}{l}{\#Actions/Item} \\
			\midrule
			ML-100k & 943          & 1,682         & 100,000         & 106.04           & 59.45            \\
                Mind    & 50,001        & 5,370         & 2,740,998        & 54.82            & 510.43           \\
                Steam   & 2,567,538      & 32,135        & 3,043,146        & 1.19             & 94.70    \\
			% Wiki & 1,070 & 4,911 & 18,778 & 17.55  & 3.82  \\
			\bottomrule
        \end{tabular}
    }

\end{table}
\vspace{-0.45in}

\subsection{Experimental Settings}
\subsubsection{Datasets.}
The experiments were conducted on three widely-used public recommendation system datasets: 
\begin{itemize}
\item 
\textbf{ML-100k} \cite{harper2015movielens}: the MovieLens-100k movie dataset, where user ratings are considered as interactions.

\item 
\textbf{Mind} \cite{wu2020mind}: the Mind news dataset, where news reading is interactions. 

\item 
\textbf{Steam} \cite{mcauley2015image}: the Steam game dataset, where user purchases are interactions.
\end{itemize}
In stage 1 of training recommendation models, we only utilize the user\-item interaction records as our training data. Following previous works, we adopt the leave-one-out \cite{wong2015performance} approach and hold out the latest interacted item of each user as the test data. We use the item before the last one for validation and the rest for training. In stage 2 of generating explanations, we employ a range of features in our approach, including interaction records, user profiles, and item attributes. The titles of items serve as the descriptive text for each item. The statistics of the three datasets are shown in \textbf{Table \ref{tab:1}}.

\subsubsection{Basic Recommendation Models.}
In stage 1, given the dataset, we use three different recommendation models: Bayesian Personalized Ranking with Matrix Factorization (BPR-MF) \cite{rendle2012bpr},  The Self-Attentive Sequential Recommendation (SASRec) \cite{kang2018self}, Light Graph Convolutional Network (LightGCN) \cite{he2020lightgcn}  to train the recommender system. 

\subsubsection{Explanation Generators.}
In stage 2, we employ several commonly used LLMs: \textbf{LLaMA} \cite{touvron2023llama2}, \textbf{ChatGLM} \cite{ChatGLM2-6B}, \textbf{GPT-3.5} \cite{ChatGPT}, \textbf{GPT-4} \cite{GPT4} and our \textbf{LLMXRec} (our instruction tuned LLaMA) as explanation generators in the second stage of our process. It should be noted that our framework cannot be directly compared with the traditional models like \cite{geng2022recommendation,cui2022m6}, since they need to be trained on user reviews.

\vspace{-0.1in}
\subsubsection{Implementation Details.}
In the first stage of training recommendation models, in the datasets of ML-100k, Mind, and Steam, we respectively employ \textbf{BPR-MF}, \textbf{SASRec}, and \textbf{LightGCN} as recommendation models and use their default parameters for training. In the second stage, upon completion of the recommendation models, for each user, we select the Top 1 item from the recommendation list as the candidate item. We take the last $L$ items as user history (up to 50 in ML-100k, 10 in Mind and Steam), using item titles for representation. For LLMXRec, we use 240 data points for LLaMA's instruction tuning. For discriminator training, we use 1440 data points for LLaMA's instruction tuning, reserving 360 for testing. For explanation via attribute prediction, we select features from various sources. From ML-100k, we choose attributes like age, gender, and occupation. From Mind, we select item categories and user interests from past interactions. From Steam, we include item price and popularity. 

% In the second stage, upon completion of the recommendation models, for each user, we select the Top 1 item from the recommendation list as the candidate item. For the current user, we choose the last $L$ items (with $L$ up to 50 in ML-100k, 10 in Mind and Steam) as the historical interaction records. We use the title of each item to represent the items in the user's historical interaction records. For LLMXRec training, We use 240 data points to employ instruction tuning for LLaMA. For discriminator training, we use 1440 data points to employ instruction tuning for LLaMA as discriminator and leave 360 data points for test. For the attribute prediction as explanation, we selected various features from different sources. From ML-100k dataset, we selected user attributes including age, gender, and occupation. From the Mind dataset, we chose the item category and the user's interests, represented as a collection of categories from their historical interaction items. Lastly, from the Steam dataset, we included the price and popularity of the items.

\vspace{-0.1in}
\subsubsection{Evaluations Metrics}  

% The task of evaluation involves determining the performance of different LLMs in generating recommendation explanations under the condition of having the same information provided (such as only giving user interaction history and candidate items).

% \textbf{(1) Win Ratio and Ranking Order}

% The metrics for automatic evaluation on explanation are the win ratio and ranking order.

% For the\textit{ win ratio}, suppose there are n LLMs, then one instruction would produce n explanations. We evaluate these in pairs, generating a total of $(^2_n)$ evaluation results, where we assign +1 score to the winner and +0 score to the loser. Finally, we aggregate the different LLMs based on their accumulated scores and compute win ratio. 

% For the \textit{ranking order}, we statistically analyze the quality of $(^2_n)$ pairs of explanations generated from a single instruction, resulting in a ranking of scores for n LLMs for that instruction. We then calculate the average ranking order across all instructions in the dataset to obtain an overall mean ranking order.
\noindent
\begin{itemize}
\item  
\textbf{Win Ratio and Ranking Order.}
The metrics for Automatic Evaluation with Fine-tuned LLMs as Discriminator in Section \ref{sect:3.4.1} are the win ratio (WR) and ranking order (RO).

% For win ratio (WR), we denote the number of LLMs as $n$. Given a single instruction, these $n$ LLMs would generate $n$ explanations. We evaluate these explanations in pairs, generating a total of $\binom{n}{2}$ evaluation results. In each pair, we assign a score of 1 to the better explanation (the winner) and a score of 0 to the other (the loser). The win ratio (WR) for a particular LLM is then computed as:

For WR, with $n$ LLMs and one instruction, they generate $n$ explanations. We pair and evaluate these, resulting in $\binom{n}{2}$ comparisons. Each pair's better explanation scores 1, the other 0. The WR for an LLM is:

\begin{equation}
WR = \frac{\text{Sum of scores for the LLM}}{\text{Total numbers of comparisons}}
\end{equation}

\noindent
% For ranking order (RO), we statistically analyze the quality of $\binom{n}{2}$ pairs of explanations generated from a single instruction, resulting in a ranking of scores for $n$ LLMs for that instruction. For each instruction, we rank the LLMs based on their scores and then calculate the average ranking order across all instructions in the dataset. The overall mean ranking order (RO) is computed as:

For ranking order (RO), we assess the quality of $\binom{n}{2}$ explanation pairs per instruction, ranking $n$ LLMs by scores for that instruction. We then average these rankings across all instructions. The mean RO is:

\begin{equation}
RO = \frac{1}{N}\sum_{i=1}^{N} \text{Rank}_i
\end{equation}
\noindent
where $N$ is the instruction count and $\text{Rank}_i$ is an LLM's rank for the $i$-th instruction. Conversely, \textit{ranking order} constitutes a list-wise comparison, providing a comprehensive view of the relative performance of multiple models in a ranked order based on their scores.

\item 
\textbf{Human Rating Score. } 
The metrics for Manual Evaluation with Scoring Explanation in Section \ref{sect:3.4.2} is the human rating score encompassing three aspects: Reasonability, Attractiveness, and Redundancy. For each of these aspects, the scores span a range from a minimum of 1 to a maximum of 10.
\item 
\textbf{Prediction Accuracy. }
The evaluation metric for Local Evaluation with Attribute Prediction in Section \ref{sect:3.4.3} is prediction accuracy. We input instructions for predicting attributes into LLMs, prompting the LLMs to infer required attributes, and calculate the accuracy of these predictions.

\end{itemize}

\subsection{Analysis on Explanation Generator}
\subsubsection{Overall Performance.}

% In order to compare the quality of recommendation explanations between the instruction-tuned LLMXRec and other baseline LLMs, we conducted experiments on three datasets and three recommendation models (1000 users for each datasets and each recommendation models), utilizing our tuned LLaMA as a discriminator for explanation comparison. The win ratio and ranking order of explanations generated by different LLMs in our tuned discriminator are shown in \textbf{Table \ref{tab:2}}. We found that the instruction-tuned LLMXRec outperforms GPT-4 in terms of the quality of explanations, followed by GPT3.5. The performance of LLaMA and ChatGLM is nearly identical, trailing behind the others. The win ratio of LLMXRec substantially improves across three datasets compared to the un-tuned LLaMA. For instance, when using the BPR-MF recommendation model on the ML-100k dataset, the win ratio increased by nearly 70\%. It is a significant improvement from the average ranking of the original LLaMA, which was around fourth place, to the first place. Moreover, the win ratio and ranking order of LLMXRec surpassed all un-tuned LLMs, demonstrating superior performance on the Mind and Steam datasets compared to the state-of-the-art model GPT-4. These results indicate that instruction tuning can significantly enhance the explainability of recommendations generated by LLMs.

To assess recommendation explanation quality, we tested the instruction-tuned LLMXRec against baseline LLMs on three datasets and  three recommendation models (1000 users each), utilizing our tuned LLaMA as a discriminator for explanation comparison, detailed in \textbf{Table \ref{tab:2}}. LLMXRec outdid GPT-4 and GPT3.5 in explanation quality, with LLaMA and ChatGLM closely matching but lower. LLMXRec's win ratio notably rose across datasets versus untuned LLaMA; for BPR-MF on ML-100k, it surged by 70\%. This leap from LLaMA's average fourth to first place, and LLMXRec's win ratio and ranking order topping all untuned LLMs, show its edge on Mind and Steam datasets over GPT-4, proving instruction tuning's potential to boost LLM-generated recommendation explainability.

\vspace{-0.33in}
{
\setlength{\tabcolsep}{2.7pt}
\begin{table*}[h!]
  \caption{LLMs' win ratio and ranking order of different recommendation models.  $\blacktriangle$\% denotes the relative improvement of win ratio and average ranking order over the best SOTA LLM.}
  \tabcolsep=0.2cm
  \renewcommand{\arraystretch}{1}
  % \vspace{-0.15in}
  \label{tab:2}
  \resizebox{1\columnwidth}{!}{
  \begin{tabular}{c|c|ccc|ccc}
    \hline
    \multirow{2}*{Rec Model} & \multirow{2}*{LLM} &\multicolumn{3}{c|}{Win Ratio}   & \multicolumn{3}{c}{Avg Ranking}     \\  \cline{3-5} \cline{6-8}
        &        & ML-100k    & Mind  & Steam    & ML-100k    & Mind  & Steam    \\
        \hline
        \multirow{6}*{BPR-MF}
        &LLaMA     & 0.176  & 0.062   & 0.229   & 4.256   & 4.750   & 4.083     \\
        &ChatGLM   & 0.084  & 0.189   & 0.025   & 4.611   & 4.242   & 4.902     \\
        &GPT3.5     & 0.515  & 0.501   & 0.496   & 2.911   & 2.994   & 3.015     \\
        &GPT4       & \underline{0.853}  & \underline{0.757}   & \underline{0.755}   & \underline{1.571}   & \underline{1.972}   & \underline{1.980}     \\
        &\textbf{LLMXRec} & \textbf{0.872}  & \textbf{0.993}   & \textbf{0.995}   & \textbf{1.488}   & \textbf{1.028}   & \textbf{1.020}     \\
        &$\blacktriangle$\%  & 2.23\% & 31.18\% & 31.79\% & 5.28\%  & 47.87\% & 48.48\%   \\
        \hline
        \multirow{6}*{Sasrec}
        &LLaMA     & 0.432  & 0.061   & 0.228   & 3.236   & 4.748   & 4.088     \\
        &ChatGLM   & 0.244  & 0.190   & 0.024   & 3.984   & 4.233   & 4.904     \\
        &GPT3.5     & 0.083  & 0.501   & 0.499   & 4.623   & 2.993   & 3.005     \\
        &GPT4       & \underline{0.842}  & \underline{0.755}   & \underline{0.760}   & \underline{1.624}   & \underline{1.976}   & \underline{1.959}     \\
        &\textbf{LLMXRec} & \textbf{0.899}  & \textbf{0.994}   & \textbf{0.990}   & \textbf{1.396}   & \textbf{1.021}   & \textbf{1.041 }    \\
        &$\blacktriangle$\%  & 6.74\% & 31.62\% & 30.19\% & 14.07\% & 48.33\% & 46.86\%   \\
        \hline
        \multirow{6}*{LightGCN}
        &LLaMA     & 0.234  & 0.185   & 0.239   & 4.035   & 4.258   & 4.044     \\
        &ChatGLM   & 0.026  & 0.065   & 0.016   & 4.857   & 4.737   & 4.937     \\
        &GPT3.5     & 0.512  & 0.501   & 0.495   & 2.930   & 2.995   & 3.019     \\
        &GPT4       & \underline{0.854}  & \underline{0.757}   & \underline{0.756}   & \underline{1.572}   & \underline{1.970}   & \underline{1.978}     \\
        &\textbf{LLMXRec} & \textbf{0.873}  & \textbf{0.992}   & \textbf{0.995}   & \textbf{1.489}   & \textbf{1.033}   & \textbf{1.022}     \\
        &$\blacktriangle$\%  & 2.17\% & 30.93\% & 31.63\% & 5.25\%  & 47.56\% & 48.33\%  \\ 

  \hline
\end{tabular}
}
\vspace{-0.3in}
\end{table*}
}
\vspace{-0.17in}

\subsubsection{Human Evaluation.}
% In order to investigate the human intuitive perception of generated explanations, we employed five experienced evaluators. We extracted 500 test cases each from each three datasets, and tasked the evaluators with assessing the reasonableness, attractiveness, and redundancy level of the generated explanations. Human rating for the recommendation explanation generated by various LLMs is displayed in \textbf{Figure \ref{fig:3}}, incorporating three aspects within the absolute scoring framework. 

% LLMXRec significantly improves the reasonability and redundancy of generated explanations at the cost of some attractiveness. For example, on the Steam dataset, it increases reasonability and redundancy scores by about 10\% compared to the original LLaMA, but reduces attractiveness by around 2\% on the Mind dataset. This is due to the model being trained to generate concise, rational explanations, which may lead to a loss in user appeal. Meanwhile, ML-100k dataset's scores are slightly higher across all datasets, likely because its movie domain data included during LLMs' pre-training enhances the quality of its explanations, a benefit absent in the Mind and Steam datasets.

\begin{figure*}
  \centering
  \includegraphics[width=1\linewidth]{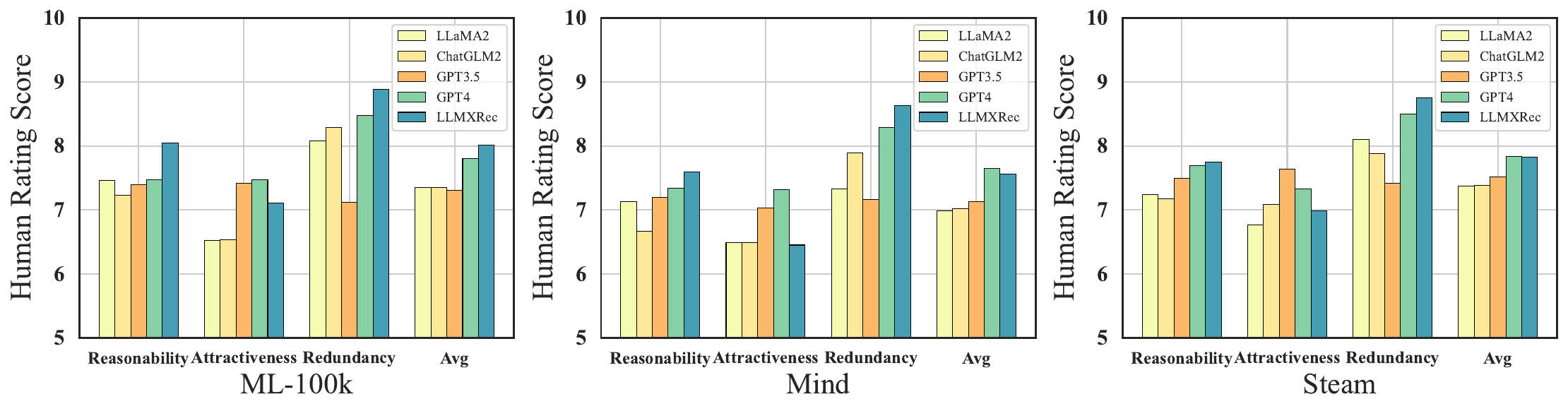}
  \vspace{-0.2in}
  \caption{Human evaluations of different LLMs}
  \label{fig:3}
  \vspace{-0.12in}
\end{figure*}

We assessed human intuitive perception of explanation generation by enlisting five expert evaluators to review 500 test cases from three datasets, judging the explanations’ reasonableness, attractiveness, and redundancy. \textbf{Figure \ref{fig:3}} shows human ratings for LLM-generated recommendation explanations, using an absolute scoring system across three datasets.

LLMXRec notably improves reasonability and reduces redundancy in explanations compared to LLaMA, with a minor drop in attractiveness. For instance, on the Steam dataset, LLMXRec's reasonability and redundancy improved by roughly 10\%, while attractiveness dipped by about 2\% on the Mind dataset, a trade-off from focusing on brevity and logic in explanations, potentially diminishing user engagement. Conversely, the ML-100k dataset scored modestly better overall, suggesting its movie-related pre-training data for LLMs enhances explanation quality, an advantage not present in Mind and Steam datasets.

% The explanations generated by LLMXRec demonstrate substantial improvements in terms of rationality and redundancy. However, this might lead to a decrease in attractiveness. For instance, the explanations produced by LLMXRec display approximately a 10\% increase in scores for rationality and redundancy on the Steam dataset, compared to the original LLaMA. Nonetheless, on the Mind dataset, the attractiveness score has decreased by about 2\% post fine-tuning. This can be attributed to the high-quality annotated explanation data, which trains the model to generate rational explanations within a word limit, thereby enhancing rationality and reducing redundancy. Simultaneously, instruction tuning may cause the model's output to conform to a specific format, which could diminish the attractiveness of the generated explanations to users.

% The average scores across all metrics for the ML-100k dataset are slightly higher than those for the Mind and Steam datasets. We believe that this is due to the inclusion of movie data in the ML-100k dataset during the pre-training of the LLMs. This exposure enables the LLMs to possess inherent knowledge about movies, thereby generating higher quality explanations, an advantage that is lacking in the Mind and Steam datasets.

\vspace{-0.2in}
\vspace{-0.05in}

\subsubsection{Local Explanation Performance.}

\vspace{-0.7in}
\begin{table*}
  \caption{The accuracy of LLMs in inferring unknown attributes based on existing ones. Age assesses if the predicted and actual ages are within a 5-year range; user interest checks if predicted interests align with categories of past interactions.}
  \vspace{0.02in}
  \label{tab:3}
  \tabcolsep=0.1cm
  \resizebox{\columnwidth}{!}{
  \begin{tabular}{c|c|ccc|cc|cc}
    \hline 
    \multirow{2}*{Rec Model} & \multirow{2}*{LLM}  &\multicolumn{3}{c|}{ML-100k}   & \multicolumn{2}{c|}{Mind}  & \multicolumn{2}{c}{Steam} \\  \cline{3-5} \cline{6-7} \cline{8-9}
    &               & Occupation            & Age          & Gender           & Item Category            & User Interest          & Popularity           & Price          \\
    \hline 
    \multirow{5}*{BPR-MF}
    &Random     & 0.016 & 0.141 & 0.500 & 0.125 & 0.360 & 0.210 & 0.153 \\
    &LLaMA     & 0.024 & 0.313 & \underline{0.743} & 0.330 & 0.642 & \underline{0.683} & \underline{0.642} \\
    &ChatGLM   & \underline{0.064} & \underline{0.375} & \textbf{0.784} & \underline{0.343} & \underline{0.673} & 0.437 & 0.233 \\
    &\textbf{LLMXRec} & \textbf{0.127} & \textbf{0.386} & 0.745 & \textbf{0.415} & \textbf{0.751} & \textbf{0.944} & \textbf{0.984} \\
    &$\blacktriangle$\% &98.44\%  & 2.93\%  & -4.97\%  & 20.99\% & 11.59\% & 38.21\% & 53.27\% \\

    \hline
    \multirow{5}*{SASRec}
    &Random     & 0.014 & 0.145 & 0.510 & 0.122 & 0.362 & 0.204 & 0.156 \\
    &LLaMA     & 0.026 & 0.294 & \underline{0.720} & 0.273 & \underline{0.684} & \underline{0.752} & \underline{0.645} \\
    &ChatGLM   & \underline{0.057} & \underline{0.385} & \textbf{0.752} & \underline{0.310} & 0.632 & 0.403 & 0.454 \\
    &\textbf{LLMXRec} & \textbf{0.114} & \textbf{0.479} & 0.671 & \textbf{0.454} & \textbf{0.771} & \textbf{0.743} & \textbf{0.953} \\
    &$\blacktriangle$\% & 100.00\% & 24.42\% & -10.77\% & 46.45\% & 12.72\% & -1.20\% & 47.75\% \\
    \hline
    \multirow{5}*{LightGCN}
    &Random     & 0.017 & 0.143 & 0.502 & 0.120 & 0.367 & 0.202 & 0.152 \\
    &LLaMA     & 0.036 & 0.302 & \underline{0.731} & 0.293 & \underline{0.650} & \underline{0.702} & \underline{0.487} \\
    &ChatGLM   & \underline{0.054} & \underline{0.342} & \textbf{0.767} & \underline{0.324} & 0.594 & 0.460 & 0.216 \\
    &\textbf{LLMXRec} & \textbf{0.125} & \textbf{0.447} & 0.724 & \textbf{0.424} & \textbf{0.782} & \textbf{0.990} & \textbf{0.960} \\
    &$\blacktriangle$\% & 131.48\% & 30.70\% & -5.61\%  & 30.86\% & 20.31\% & 41.03\% & 97.13\%  \\
    \hline
\end{tabular}}
\vspace{-0.2in}
\end{table*}

To further probe into the explainability capabilities of LLMs, we opt to input various attributes into LLMs and generate explanations.  The experimental results are presented in the \textbf{Table \ref{tab:3}}. Given the exceptional alignment of GPT3.5 and GPT4 with human, and their inability to generate uncertain answers, we do not include these two LLMs in our comparison.

% \begin{figure}
%   \centering
%   \includesvg[width=0.85\linewidth]{image/history}
%   % \vspace{-0.2in}
%   \caption{Win ratio of different historical interaction lengths (compare to length 10)}
%   \label{fig:4}
%   % \vspace{-0.2in}
% \end{figure}

We find that LLMs can extract context and predict attributes with accuracy surpassing random prediction, notably in the ML-100k and Mind datasets, likely due to their richer information. The prediction accuracy of "popularity" is linked to the recommendation model, and LLMs' ability to predict user features also outperforms random prediction. 

After instruction tuning, LLMXRec's inferential capabilities significantly improve. Fine-tuning with instruction data of attributes led to increased accuracy in predicting the "Occupation" attribute in the ML-100k dataset and "popularity" and "price" attributes in the Steam dataset. This is because instruction tuning bolsters LLMs' ability to infer unknown attributes from user characteristics and to reconstruct user, item, and global attributes from limited information, enabling more reasonable explanations.

\subsection{More Analysis on Explanation Quality}
To enhance the quality of explanations, our research has focused on two aspects: (1) the prompt level, which involves determining how to modify input instructions to generate improved explanations; and (2) the instruction tuning level, which examines how to fine-tune the model to produce superior explanations.

\vspace{-0.2in}
\begin{figure}[htbp]
  \centering
  % 第一张子图
  \begin{subfigure}[b]{0.45\textwidth}
    \includegraphics[width=\textwidth]{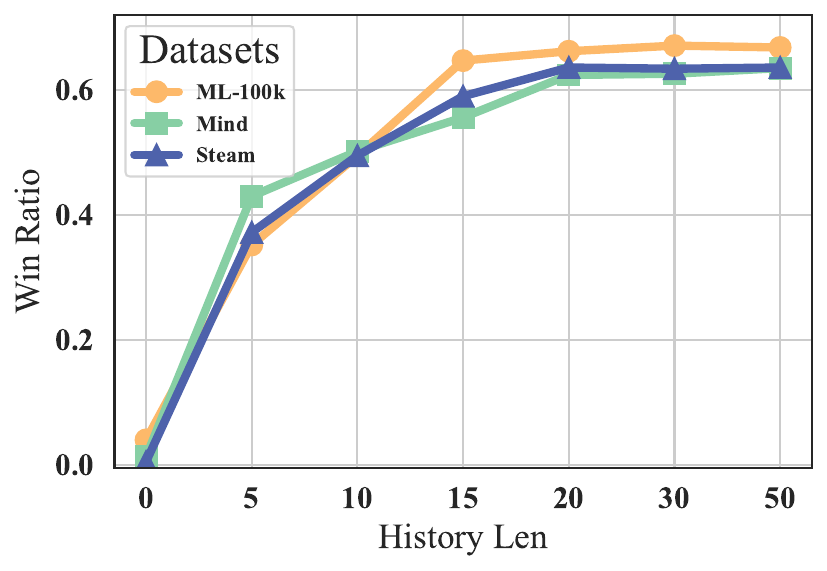}
    \caption{}
    % \vspace{-0.1in}
    \label{fig:5.a}
  \end{subfigure}
  \hfill % 添加一些水平间距
  % 第二张子图
  \begin{subfigure}[b]{0.45\textwidth}
    \includegraphics[width=\textwidth]{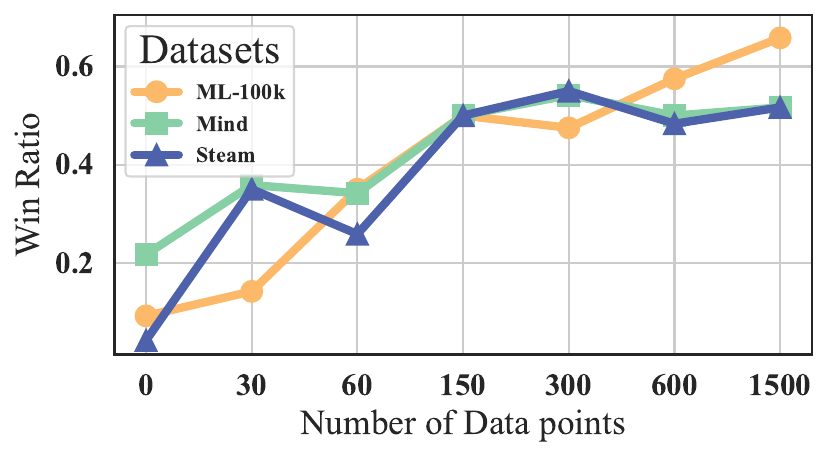}
    \caption{}
    % \vspace{-0.1in}
    \label{fig:5.b}
  \end{subfigure}
  \vspace{-0.1in}
  \caption{(a) is Win ratio of different historical interaction lengths (compare to length 10). (b) is Win ratio of fine-tuning LLaMA with different data sizes (compared to fine-tuning using 150 data points)}
  \label{fig:test}
\end{figure}
\vspace{-0.45in}

\subsubsection{Prompt Design.}
We aim to investigate the impact of different features and properties input into LLMs in generating explanations.

 % We first explored the effect of the length of the user's input history sequence. We input history sequences of different lengths into LLaMA to generate explanations and compared them with explanations generated from an input sequence length of 10. The results are displayed in \textbf{Figure \ref{fig:4}}. All three datasets exhibit the same trend: the longer the input history sequence, the higher thewin ratio. However, when the sequence lengthens to a certain extent, the quality of the explanation no longer improves. For instance, on the Steam and Mind datasets, thewin ratio barely changes when the length of the input history sequence increases to 20. We believe that the LLMs have fully extracted user interests from a sequence of length 20, and further increasing the sequence length has almost no impact on the extraction of user interests by LLMs, resulting in a non-increasingwin ratio.

 % Our study on the effect of user's input history length in \textbf{Figure \ref{fig:5.a}} reveals a trend across all datasets: longer sequences lead to higher win ratios. However, explanation quality plateaus after a certain length. For instance, on the Steam and Mind datasets, increasing the sequence length to 20 doesn't significantly enhance the win ratio. This suggests that LLMs can fully extract user interests from a sequence of this length, and further lengthening has minimal impact on user interest extraction, hence the stagnant win ratio.

In \textbf{Figure \ref{fig:5.a}}, our analysis shows longer input histories increase win ratios across datasets, but plateau beyond a certain length. For example, in Steam and Mind datasets, a sequence length of 20 doesn't markedly improve win ratios, indicating LLMs capture user interests adequately by this length, with additional length having little effect on interest extraction and thus, stable win ratios.
% Furthermore, we compare the quality of explanations generated by inputting the same history sequence length but adding different side information. The results are shown in \textbf{Figure \ref{fig:3}}. Adding user profiles, such as age, gender, occupation, etc., can improve the quality of the explanation. For example, in the ML-100k dataset, the win ratio of explanations generated with user profiles compared to those without reached 85.1\%. Similarly, adding item attributes, such as item categories, to the history sequence or recommended items can also increase the win ratio of the explanation. This indicates side information has a positive effect on the quality of the explanation.

% The comparison of explanations in \textbf{Figure \ref{fig:4}}, using the same history sequence length but varying side information, reveals that adding user profiles or item attributes improves explanation quality. For instance, in the ML-100k dataset, explanations with user profiles had an 85.1\% win ratio over those without. Similarly, adding item categories to the history sequence or recommended items also increased the win ratio. This suggests that side information positively impacts explanation quality.

\begin{figure*}
  \centering
  \includegraphics[width=0.95\linewidth]{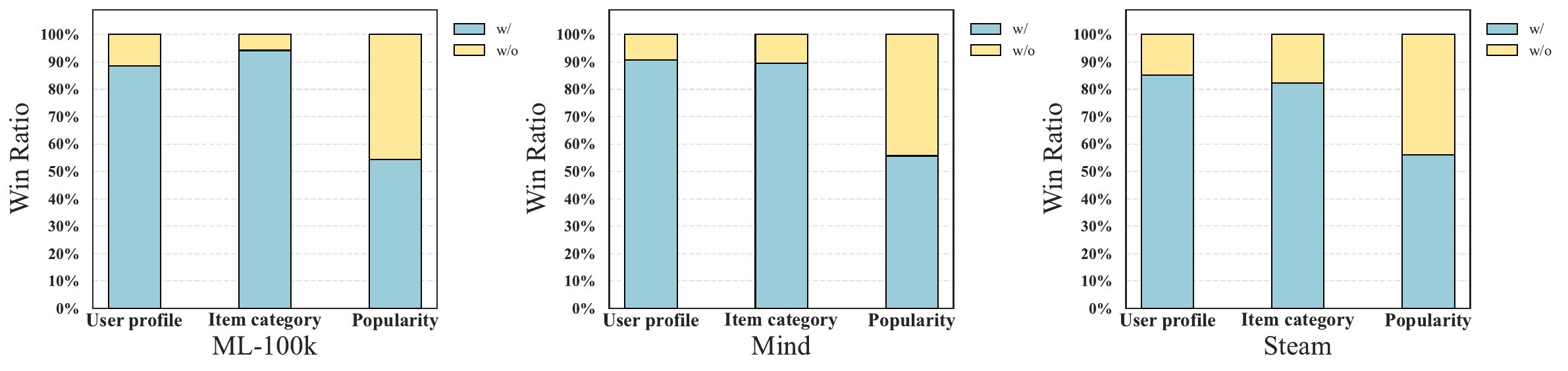}
  \vspace{-0.12in}
  \caption{Comparison of win ratio with and without added
attributes}
  \label{fig:4}
  \vspace{-0.3in}
\end{figure*}

The comparison of explanations in \textbf{Figure \ref{fig:4}}, using the same history sequence length but varying side information, reveals that adding user profiles or item attributes improves explanation quality. For example, in ML-100k, explanations including user profiles achieved an 85.1\% win ratio over those without. Likewise, incorporating item categories into history or recommended items raised win ratios, indicating side information enhances explanation quality.
% \begin{figure}
%   \centering
%   \includesvg[width=0.8\linewidth]{image/diff_len}
%   \vspace{-0.22in}
%   \caption{Win ratio of fine-tuning LLaMA with different data sizes (compared to fine-tuning using 150 data points)}
%   \label{fig:6}
%   \vspace{-0.16in}
% \end{figure}

\vspace{-0.15in}

\subsubsection{Instruction Tuning LLMs with Varying Amounts of Data.}
 We investigate the impact of varying quantities of high-quality human-annotated data used to tune LLMXRec in generating explanations. We compared LLaMA fine-tuned with varying amounts of data to a baseline LLMXRec model trained with 150 human-annotated explanations. Results in the\textbf{ Figure \ref{fig:5.b}} show that as we increase the fine-tuning data, the quality of explanations improves, reflected in a higher win ratio. For Mind and Steam datasets, the win ratio plateaus after 300 data points, indicating optimal quality. Yet, with the ML-100k dataset, the win ratio keeps rising with more data. This suggests that while more high-quality instruction data improves explanation quality, there's a saturation point beyond which additional data doesn't significantly enhance quality.

 % We use LLMXRec tuned with 150 human-annotated explanation data as baseline, apply different quantities of data to fine-tune LLaMA with instructions, and compare it with the baseline using the trained discriminator. The results are displayed in the\textbf{ Figure \ref{fig:5}}, where the win ratio of 0.5 indicates that the quality of explanations is equivalent to the baseline. As the quantity of fine-tuning data increases, the win ratio of the explanations produced by the LLM also increases. For Mind and Steam datasets, LLMXRec's win ratio plateaus after 300 data points, suggesting optimal explanation quality at this point. However, with the ML-100k dataset, the win ratio continues to rise with more data. The overall trend indicates that fine-tuning LLMs with more high-quality instruction data can produce better quality explanations, but not linearly, showing a saturation point where extra data doesn't significantly boost quality.

\subsection{Case Study}
\vspace{-0.2in}
\begin{figure}
  \centering
  \includegraphics[width=0.99\linewidth]{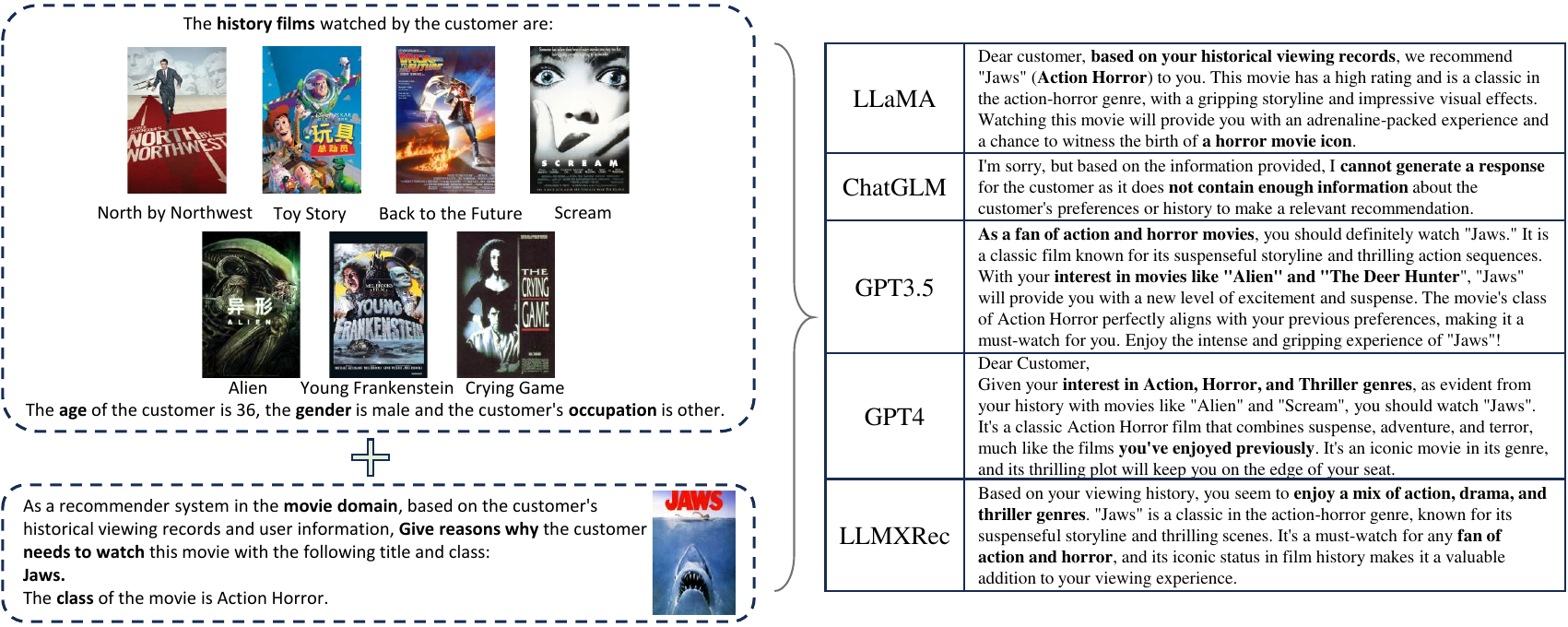}
  \vspace{-0.1in}
  \caption{Explanation of LLMs with the above instruction}
  \label{fig:7}
  \vspace{-0.2in}
\end{figure}

% In \textbf{Figure \ref{fig:7}}, we display cases of explanations from LLMXRec and other LLMs, using the same instruction for all to demonstrate the correlation between LLMs and generated explanations. Our LLMXRec excels at extracting user interests and linking them to recommended item traits. Some LLMs, like ChatGLM, may produce unsatisfactory responses due to constraints, lacking sufficient inference information. Utilizing instruction tuning, LLMXRec reduces unexpected errors and provides more valuable explanations, surpassing other LLMs in rationality and redundancy.

\textbf{Figure \ref{fig:7}} compares explanations from LLMXRec and other LLMs, illustrating their correlation with generated content. LLMXRec outperforms in aligning user interests with item features. Certain LLMs, e.g., ChatGLM, might yield subpar results from limitations in inference data. Through instruction tuning, LLMXRec minimizes errors and delivers superior explanations, leading in rationality and conciseness over other LLMs.

% We also conducted a group case study on the ML-100k dataset, investigating the relationship between the word count of explanations generated by LLMs and the gender of the users. The results are presented in \textbf{Table \ref{tab:4}}. We observed a bias in the LLMs' generation of explanations, wherein all LLMs tended to generate more explanations for female users. This bias was least pronounced in GPT-4. Our LLMXRec, after instruction tuning, was able to mitigate this bias, exhibiting a performance second only to GPT-4.

Our group case study on the ML-100k dataset examined LLM-generated explanation length versus user gender, with findings in \textbf{Table \ref{tab:4}}. All LLMs showed a tendency to produce longer explanations for female users, with GPT-4 having the least bias. By instruction tuning, our LLMXRec reduced this bias, ranking just behind GPT-4.

\vspace{-0.2in}
\begin{table}
  \caption{Word counts of explanations with gender}
  \renewcommand{\arraystretch}{1}
  \tabcolsep=0.5cm
  % \vspace{-0.15in}
  \label{tab:4}
  \resizebox{1\columnwidth}{!}{
  \begin{tabular}{c|ccccc}
    \hline
    Gender& LLaMA & ChatGLM & GPT3.5 & GPT4  & LLMXRec \\
    \hline       
    Male   & 79.01 & 78.95   & 80.56  & 75.56 & 69.74   \\
    Female & 85.32 & 84.58   & 86.32  & 76.56 & 72.53   \\
    \hline
    $\Delta$    & 6.31  & 5.63    & 5.76   & \textbf{1.00}  & \underline{2.79}    \\
\hline
\end{tabular}
}
% \vspace{-0.3in}
\end{table}

% \vspace{-0.1in}
\section{Conclusion}
In this paper, we proposed a two-stage framework for explainable recommendation, utilizing Large Language Models(LLMs) as explanation generators. Our framework is model-agnostic in relation to the recommendation model, offering excellent explainability, and flexibly generating personalized explanations based on the application domain. To produce higher quality and more accurate explanations, we employed instruction tuning to enhance the controllability of the LLMs. To demonstrate the effectiveness of our framework, we designed three evaluation methods, encompassing both quantitative and qualitative dimensions. The experiment results affirm the efficacy of our proposed framework.

\noindent
\textbf{Limitations and future work.}
The evaluation results prove the effectiveness of our method. However, there are indeed some limitations. Firstly, LLMs may generate technically accurate yet unhelpful or incomprehensible explanations. Secondly, the explanations produced by LLMs exhibit a certain degree of bias, such as explanation length correlating with user gender. Our future research will involve leveraging our explainable framework to enhance the accuracy of recommendation systems and utilizing the explainability of recommendations to discern the underlying causes of unsuccessful recommendations. Moreover, we are interested in the potential of LLMs in generating bias-free explanations and providing a user-friendly library of explainability for recommendation systems.

\bibliographystyle{splncs04}
\bibliography{sample-base}

\begin{thebibliography}{10}
\providecommand{\url}[1]{\texttt{#1}}
\providecommand{\urlprefix}{URL }
\providecommand{\doi}[1]{https://doi.org/#1}

\bibitem{bilgic2005explaining}
Bilgic, M., Mooney, R.J.: Explaining recommendations: Satisfaction vs. promotion. In: Beyond personalization workshop, IUI. vol.~5, p.~153 (2005)

\bibitem{brown2020language}
Brown, T., Mann, B., Ryder, N., Subbiah, M., Kaplan, J.D., Dhariwal, P., Neelakantan, A., Shyam, P., Sastry, G., Askell, A., et~al.: Language models are few-shot learners. Advances in neural information processing systems  \textbf{33},  1877--1901 (2020)

\bibitem{chen2018neural}
Chen, C., Zhang, M., Liu, Y., Ma, S.: Neural attentional rating regression with review-level explanations. In: Proceedings of the 2018 world wide web conference. pp. 1583--1592 (2018)

\bibitem{chen2018visually}
Chen, X., Zhang, Y., Xu, H., Cao, Y., Qin, Z., Zha, H.: Visually explainable recommendation. arXiv preprint arXiv:1801.10288  (2018)

\bibitem{cheng2022towards}
Cheng, M., Liu, Z., Liu, Q., Ge, S., Chen, E.: Towards automatic discovering of deep hybrid network architecture for sequential recommendation. In: Proceedings of the ACM Web Conference 2022. pp. 1923--1932 (2022)

\bibitem{cheng2021learning}
Cheng, M., Yuan, F., Liu, Q., Xin, X., Chen, E.: Learning transferable user representations with sequential behaviors via contrastive pre-training. In: 2021 IEEE International Conference on Data Mining (ICDM). pp. 51--60. IEEE (2021)

\bibitem{cui2022m6}
Cui, Z., Ma, J., Zhou, C., Zhou, J., Yang, H.: M6-rec: Generative pretrained language models are open-ended recommender systems. arXiv preprint arXiv:2205.08084  (2022)

\bibitem{diao2014jointly}
Diao, Q., Qiu, M., Wu, C.Y., Smola, A.J., Jiang, J., Wang, C.: Jointly modeling aspects, ratings and sentiments for movie recommendation (jmars). In: Proceedings of the 20th ACM SIGKDD international conference on Knowledge discovery and data mining. pp. 193--202 (2014)

\bibitem{gao2023chat}
Gao, Y., Sheng, T., Xiang, Y., Xiong, Y., Wang, H., Zhang, J.: Chat-rec: Towards interactive and explainable llms-augmented recommender system. arXiv preprint arXiv:2303.14524  (2023)

\bibitem{geng2022recommendation}
Geng, S., Liu, S., Fu, Z., Ge, Y., Zhang, Y.: Recommendation as language processing (rlp): A unified pretrain, personalized prompt \& predict paradigm (p5). In: Proceedings of the 16th ACM Conference on Recommender Systems. pp. 299--315 (2022)

\bibitem{harper2015movielens}
Harper, F.M., Konstan, J.A.: The movielens datasets: History and context. Acm transactions on interactive intelligent systems (tiis)  \textbf{5}(4),  1--19 (2015)

\bibitem{he2020lightgcn}
He, X., Deng, K., Wang, X., Li, Y., Zhang, Y., Wang, M.: Lightgcn: Simplifying and powering graph convolution network for recommendation. In: Proceedings of the 43rd International ACM SIGIR conference on research and development in Information Retrieval. pp. 639--648 (2020)

\bibitem{kang2018self}
Kang, W.C., McAuley, J.: Self-attentive sequential recommendation. In: 2018 IEEE international conference on data mining (ICDM). pp. 197--206. IEEE (2018)

\bibitem{Alpaca}
tatsu lab: Alpaca (2023), \url{https://github.com/tatsu-lab/stanford\_alpaca}

\bibitem{mcauley2013hidden}
McAuley, J., Leskovec, J.: Hidden factors and hidden topics: understanding rating dimensions with review text. In: Proceedings of the 7th ACM conference on Recommender systems. pp. 165--172 (2013)

\bibitem{mcauley2015image}
McAuley, J., Targett, C., Shi, Q., Van Den~Hengel, A.: Image-based recommendations on styles and substitutes. In: Proceedings of the 38th international ACM SIGIR conference on research and development in information retrieval. pp. 43--52 (2015)

\bibitem{ChatGPT}
OpenAI: Chatgpt (mar 14 version)  (2023), \url{https://chat.openai.com/chat}

\bibitem{GPT4}
OpenAI: Gpt-4 technical report. CoRR abs/2303.08774  (2023)

\bibitem{radford2019language}
Radford, A., Wu, J., Child, R., Luan, D., Amodei, D., Sutskever, I., et~al.: Language models are unsupervised multitask learners. OpenAI blog  \textbf{1}(8), ~9 (2019)

\bibitem{rendle2012bpr}
Rendle, S., Freudenthaler, C., Gantner, Z., Schmidt-Thieme, L.: Bpr: Bayesian personalized ranking from implicit feedback. arXiv preprint arXiv:1205.2618  (2012)

\bibitem{resnick1997recommender}
Resnick, P., Varian, H.R.: Recommender systems. Communications of the ACM  \textbf{40}(3),  56--58 (1997)

\bibitem{sharma2013social}
Sharma, A., Cosley, D.: Do social explanations work? studying and modeling the effects of social explanations in recommender systems. In: Proceedings of the 22nd international conference on World Wide Web. pp. 1133--1144 (2013)

\bibitem{ChatGLM2-6B}
THUDM: Chatglm2-6b (2023), \url{https://github.com/THUDM/ChatGLM2-6B}

\bibitem{tintarev2010designing}
Tintarev, N., Masthoff, J.: Designing and evaluating explanations for recommender systems. In: Recommender systems handbook, pp. 479--510. Springer (2010)

\bibitem{touvron2023llama2}
Touvron, H., Martin, L., Stone, K., Albert, P., Almahairi, A., Babaei, Y., Bashlykov, N., Batra, S., Bhargava, P., Bhosale, S., et~al.: Llama 2: Open foundation and fine-tuned chat models. arXiv preprint arXiv:2307.09288  (2023)

\bibitem{wang2018reinforcement}
Wang, X., Chen, Y., Yang, J., Wu, L., Wu, Z., Xie, X.: A reinforcement learning framework for explainable recommendation. In: 2018 IEEE international conference on data mining (ICDM). pp. 587--596. IEEE (2018)

\bibitem{wei2022chain}
Wei, J., Wang, X., Schuurmans, D., Bosma, M., Xia, F., Chi, E., Le, Q.V., Zhou, D., et~al.: Chain-of-thought prompting elicits reasoning in large language models. Advances in Neural Information Processing Systems  \textbf{35},  24824--24837 (2022)

\bibitem{wong2015performance}
Wong, T.T.: Performance evaluation of classification algorithms by k-fold and leave-one-out cross validation. Pattern recognition  \textbf{48}(9),  2839--2846 (2015)

\bibitem{wu2020mind}
Wu, F., Qiao, Y., Chen, J.H., Wu, C., Qi, T., Lian, J., Liu, D., Xie, X., Gao, J., Wu, W., et~al.: Mind: A large-scale dataset for news recommendation. In: Proceedings of the 58th Annual Meeting of the Association for Computational Linguistics. pp. 3597--3606 (2020)

\bibitem{zhang2020explainable}
Zhang, Y., Chen, X., et~al.: Explainable recommendation: A survey and new perspectives. Foundations and Trends{\textregistered} in Information Retrieval  \textbf{14}(1),  1--101 (2020)

\bibitem{zhang2014explicit}
Zhang, Y., Lai, G., Zhang, M., Zhang, Y., Liu, Y., Ma, S.: Explicit factor models for explainable recommendation based on phrase-level sentiment analysis. In: Proceedings of the 37th international ACM SIGIR conference on Research \& development in information retrieval. pp. 83--92 (2014)

\end{thebibliography}

\end{document}